\documentclass{IEEEtran}
\usepackage{cite}
\usepackage{amsmath,amssymb,amsfonts}
\usepackage{algorithmic}
\usepackage{graphicx}
\usepackage{textcomp}
\usepackage{color}
\def\BibTeX{{\rm B\kern-.05em{\sc i\kern-.025em b}\kern-.08em
    T\kern-.1667em\lower.7ex\hbox{E}\kern-.125emX}}
\begin{document}
\title{Slice Imputation: Intermediate Slice Interpolation for Anisotropic 3D Medical Image Segmentation}
\author{Zhaotao Wu, Jia Wei, Jiabing Wang, and Rui Li
\thanks{This work is supported in part by the Natural Science Foundation of Guangdong Province (2020A1515010717), the Fundamental Research Funds for the Central Universities (2019MS073), NSF-1850492 (to R.L.), and NSF-2045804 (to R.L.).}
\thanks{Zhaotao Wu is with the School of Computer Science and Engineering, South China University of Technology, Guangzhou, China (e-mail: csztwu@mail.scut.edu.cn).}
\thanks{Jia Wei is the corresponding author. He is with the School of Computer Science and Engineering, South China University of Technology, Guangzhou, China (e-mail: csjwei@scut.edu.cn).}
\thanks{Jiabing Wang is with the School of Computer Science and Engineering, South China University of Technology, Guangzhou, China (e-mail: jbwang@scut.edu.cn).}
\thanks{Rui Li is with the Golisano College of Computing and Information Sciences, Rochester Institute of Technology, Rochester, NY 14623, USA (e-mail: rxlics@rit.edu).}}
\maketitle

\begin{abstract}
We introduce a novel frame-interpolation-based method for slice imputation to improve segmentation accuracy for anisotropic 3D medical images, in which the number of slices and their corresponding segmentation labels {can be} increased between two consecutive slices in anisotropic 3D medical volumes. Unlike previous inter-slice imputation methods, which only focus on the smoothness in the axial direction, this study aims to improve the smoothness of the interpolated 3D medical volumes in all three directions: axial, sagittal, and coronal. The proposed multitask inter-slice imputation method, in particular, incorporates a smoothness loss function to evaluate the smoothness of the interpolated 3D medical volumes in the through-plane direction (sagittal and coronal). It not only improves the resolution of the interpolated 3D medical volumes in the through-plane direction but also transforms them into isotropic representations, which leads to better segmentation performances. Experiments on whole tumor segmentation in the brain, liver tumor segmentation, and prostate segmentation indicate that our method outperforms the competing slice imputation methods on both computed tomography and magnetic resonance images volumes in most cases.
\end{abstract}


\section{Introduction}
\label{sec:introduction}

In the field of medical image processing and analysis, medical image segmentation is a difficult but critical task. It is a crucial step in image-guided surgery, computer-aided detection, and medical data visualization \cite{b0}--\cite{b000}. Its goal is to accurately segment medical images with semantic labels so that it can provide reliable meaningful information for clinical diagnosis and pathology research. Moreover, it aims to assist physicians in making correct diagnoses. Deep-learning-based methods for medical image segmentation have been proposed in recent years and have demonstrated state-of-the-art performance \cite{b0.1}--\cite{b0.3}. 

For 3D medical image segmentation, deep-learning-based methods prefer to learn features from isotropic volume data as they can provide more anatomical details\cite{b1}. 
However, due to hardware limitations and time costs, isotropic volumes are difficult to obtain in clinical practice\cite{b2}. Anisotropic volumes are commonly available in most cases. In the through-plane directions, an anisotropic 3D volume is elongated, resulting unequal resolutions in the three dimensions \cite{b2.5}. For example, it may have high resolution (HR) in the in-plane (axial) but low resolution (LR) in the through-plane (sagittal and coronal) directions. Consequently, detailed structures in the through-plane direction are unclear, which leads to a negative impact on image analysis, visualization, and diagnosis of lesions \cite{b1}.

\begin{figure}
\centerline{\includegraphics[scale=0.8]{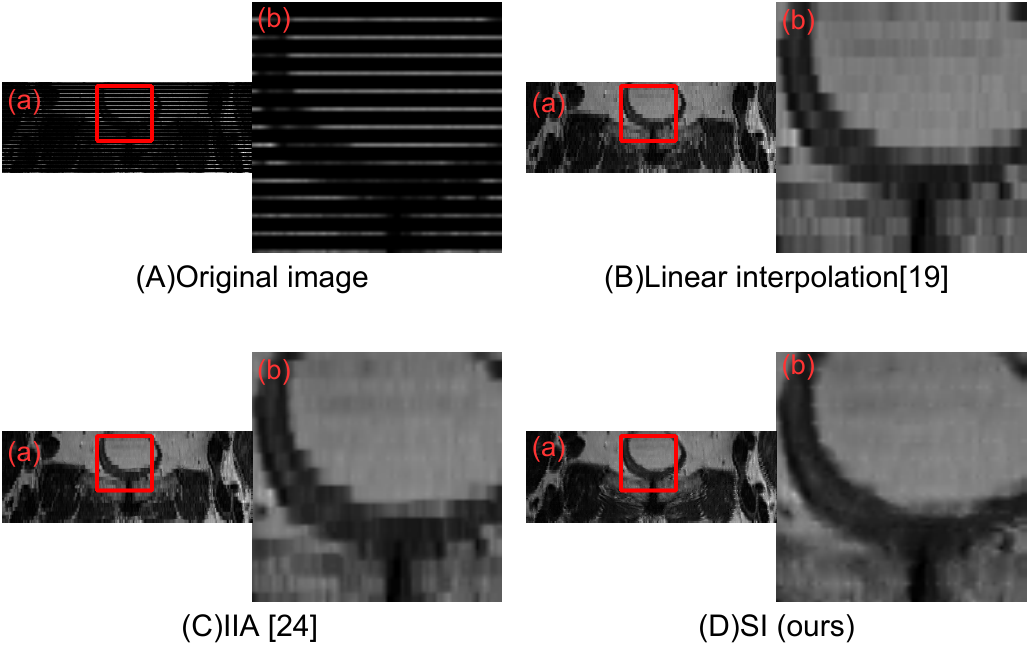}}
\caption{Synthetic slices of different methods in the through-plane direction.} 
\label{fig0}
\end{figure}

In the field of medical image segmentation, segmentation on anisotropic 3D medical volumes remains a difficult task. The main problem is that anisotropic 3D medical volumes are too sparse to adequately fit functional representations or provide sufficient fine-scale information to recover the missing details \cite{b3}. Image clarity is reduced and anatomical structures are significantly distorted between consecutive slices when we directly increase apparent volume resolution, for example, using linear interpolation (Linear) \cite{b18}, as presented in Fig.~\ref{fig0}(B).

We address the challenge described above by proposing a method to synthesize intermediate slices between consecutive slices. 3D medical volumes are continuous slice sequences in the dimension of space, similar to videos, which are continuous image sequences in the dimension of time. As a result, we propose a multiple intermediate slices interpolation method, called slice imputation (SI), to generate isotropic 3D volumes and make them suitable for segmentation, inspired by the idea of frame interpolation \cite{b28}. The main contributions of this work are as follows:
\begin{itemize}
\item To solve the anisotropy problem of 3D medical volumes for medical image segmentation, we use a frame interpolation method. The inter-slice distance of the interpolated volumes can become close to the x-y spacing distance by increasing the number of slices between every two consecutive slices, transforming them into isotropic volumes.
\item To improve slice smoothness in the through-plane direction, we introduce a smoothness loss function to evaluate the smoothness of 3D medical volumes in the through-plane direction.
\item To improve the authenticity of the interpolated slices, we employ a multitask learning mechanism. The model can use the interacting information between the classification and distinguishing tasks to generate more realistic slices by 
learning the object classifier and the local discriminator together.
\end{itemize}

This research is a significant extension of our previous work\cite{b13}. The smoothness in the in-plane direction is the only focus of the inter-slice image augmentation (IIA) method proposed in \cite{b13}. The interpolated 3D volumes tend to have blurry edges in the through-plane direction, as presented in Fig.~\ref{fig0}(C). As can be seen from Fig.~\ref{fig0}(D), the work proposed in this paper addresses the problem with significantly clearer volumes by extending IIA. 

The remainder of this paper is laid out as follows: In Section \ref{s2}, we go over some related works. The SI method and its derivation are discussed in Section \ref{s3}. We describe the implementation details in Section \ref{s4}, as well as experimental results and an analysis of the algorithm's behavior. 
{Section \ref{s4.5} performs a detailed ablation analysis of the network to validate the effect of the local and global discriminators and determine the optimal numbers of intermediate slices. }
Finally we present a conclusion in Section \ref{s5}.

\section{Related Work}
\label{s2}
\subsection{Medical image segmentation for anisotropic volumes}
In the field of medical image segmentation, deep-learning-based medical image segmentation methods have recently achieved state-of-the-art performance \cite{b18}, \cite{b14}--\cite{b17}. However, due to the anisotropy of the 3D volumes, deep-learning-based methods still fall short in 3D medical image segmentation \cite{a17}. Some recent studies focused on addressing this anisotropy problem \cite{b5}, \cite{b4}.

Lee et al. \cite{b5} proposed to avoid downsampling feature maps along the z-dimension and used convolution kernels with a particular size, which transforms the volumes into almost isotropic ones. Based on 2D and 3D vanilla U-Nets, Isensee et al. \cite{b4} proposed a robust and self-adapting framework. This method makes the volumes isotropic by first downsampling the HR axes of the anisotropic medical volumes until they match the LR axes and then upsampling the volumes to the original voxel spacing. However, directly downsampling the higher-resolution axes of the volumes may lead to information loss, which may lead to a negative impact on high-quality segmentation. On the contrary, our method, namely, SI, preserves information in the original medical volumes by increasing the slices between every two consecutive slices while transforming the volumes into isotropic ones.

\subsection{Super-Resolution algorithm}
Super-resolution (SR) methods aim to learn complex mapping relations between LR and HR images. Kim et al. \cite{b20} proposed a deep-convolutional-network based SR method. By increasing the network depth, this method significantly improves the accuracy of restoration. Although deep SR methods achieve accurate restorations of high-frequency contents, effectively training a very deep SR CNN is challenging due to the vanishing gradient problem \cite{b1}. Du et al. \cite{b1} proposed an SR reconstruction method based on residual learning with long and short skip connections. Deep networks’ vanishing gradient problem can be effectively addressed with the proposed method, which restores high-frequency details of magnetic resonance images (MRI). 
Most of the existing SR algorithms, according to Lim et al. \cite{b22_12}, treat super-resolution of different scale factors as independent problems without considering mutual relationships among different scales in SR. As a result, they proposed the EDSR, an enhanced deep SR network, that transfers knowledge from a model trained at other scales. Zhang et al. \cite{b22_13} proposed a residual channel attention network (RCAN) to obtain very deep trainable networks and adaptively learn informative channel-wise features.

In the field of SR, generative Adversarial Networks (GANs) \cite{b22_14} are also used to improve the visual quality of the generated images{, called super-resolution generative adversarial network (SRGAN) \cite{b22_14_5}. However, the hallucinated details of SRGAN are often accompanied with unpleasant artifacts. For this reason, Wang et al. \cite{b22_15} proposed an enhanced super-resolution generative adversarial network (ESRGAN). They} revisited the key components of SRGAN and improved the model by introducing the residual dense block without batch normalization as the basic network building unit.

Although the aforementioned methods perform well at reconstructing HR images from LR ones, they cannot generate corresponding segmentation labels, and some of them may change the original slices. On the contrary, SI can use bidirectional spatial transformations to generate intermediate slices and segmentation labels between every two consecutive slices. For this reason, SI can generate the HR volumes without changing the original slices, and the corresponding segmentation labels can be directly generated using the bidirectional spatial transformations. 

\subsection{Image augmentation based on spatial transformation}
A large amount of training data is critical to the success of deep learning. However, in the field of medical imaging, a lack of training data is a significant challenge. Due to issues such as a lack of cases, insufficient medical resources, and costly labeling, researchers have turned to data augmentation to better utilize existing data \cite{b21.4}, \cite{b21.5}. For medical images, data augmentation is commonly preferred using random smooth flow fields to simulate anatomical variations \cite{b14}. Although this method can reduce overfitting and improve test performance \cite{21.6}, \cite{21.7}, the selection of transformation functions and parameter settings tend to influence the improvement of performance\cite{b24}.

Data augmentation methods based on learning spatial transformations from existing data have been proposed \cite{b23}, \cite{b24}. Hauberg et al. \cite{b23} aimed to improve MNIST digit classification performance through data enhancement. It learns digit-specific spatial transformations and samples training images and transformations to create new examples. Zhao et al. \cite{b24} proposed an automatic augmentation method that has the potential to improve the performance of brain MRI segmentation. The set of spatial and appearance transformations between the labeled atlas and unlabeled volumes is modeled using learning-based registration methods. It can use unlabeled images to synthesize diverse and realistic labeled samples by capturing anatomical and imaging diversity. However, these methods cannot be directly applied to our problem scenarios. The synthetic slices of these methods cannot be interpolated into the original volumes to make them isotropic because the slice smoothness is ignored. In our previous work \cite{b13}, we proposed a method that generates synthetic inter-slice images based on frame interpolation and attention mechanism, called IIA. IIA makes use of the idea of frame interpolation to generate spatial transformation between two consecutive slices. The method can generate as many intermediate slices as needed by employing spatial transformations. However, IIA only focuses on smoothness in the in-plane direction, and it tends to perform poorly in generating 3D volumes with clear edges in the through-plane direction. SI proposes the use of a smoothness loss function that can evaluate the smoothness of 3D medical volumes in the through-plane direction to generate medical volumes with significantly clearer edges in order to improve slice smoothness in the through-plane direction.

\section{Methods}
\label{s3}
\begin{figure*}
  \centerline{\includegraphics[scale=0.35]{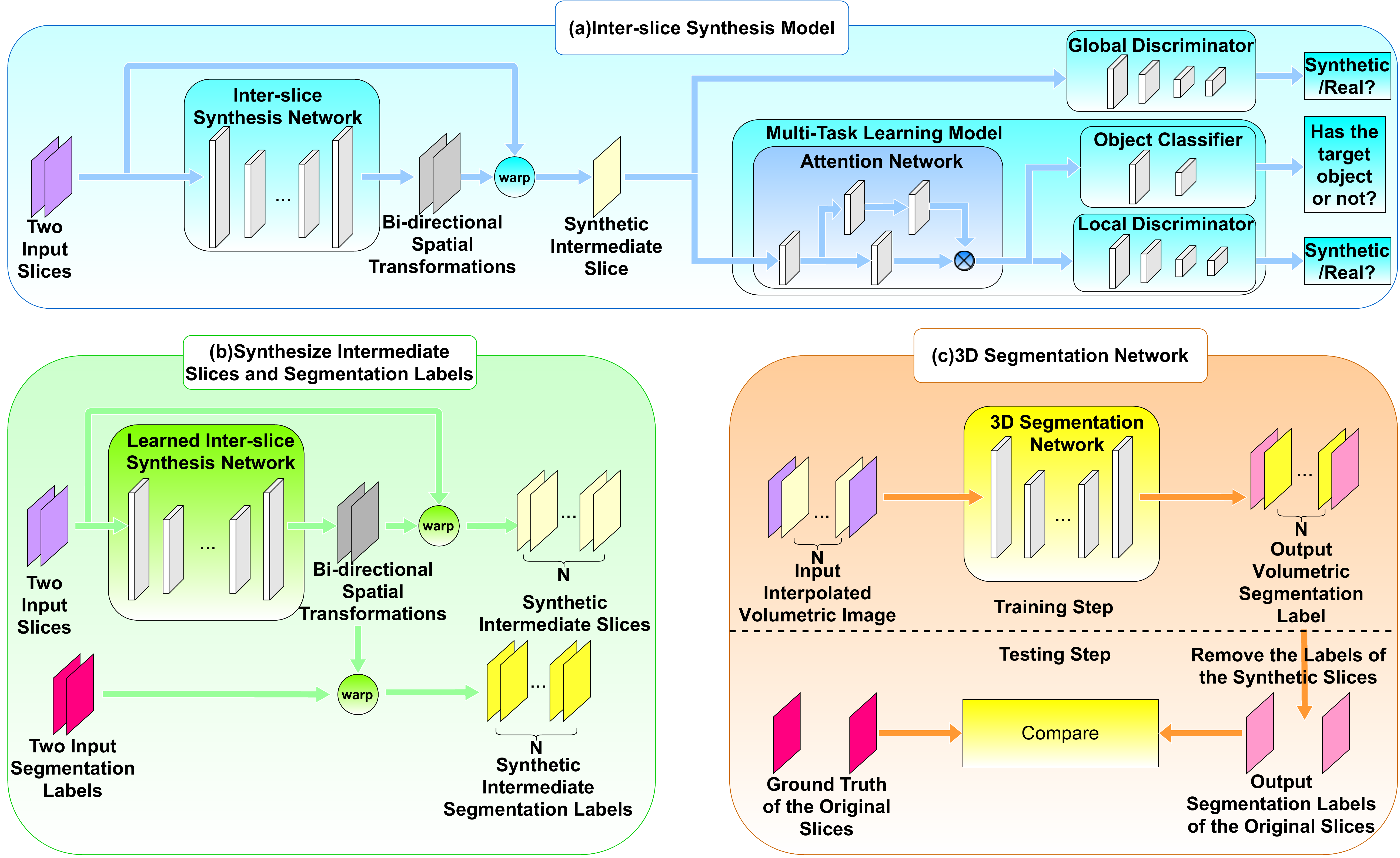}}
    \caption{An overview of the proposed method: (a) the architecture of the inter-slice synthesis model, (b) the process of synthesizing intermediate slices and their corresponding segmentation labels, and (c) the architecture of the 3D segmentation network. $N$ is the number of interpolated slices between two consecutive slices.}
  \label{fig1}
\end{figure*}

By making 3D medical volumes isotropic and clear in both the in- and through-plane directions, we proposed a method, namely, SI, to improve the 3D medical image segmentation accuracy. Fig.~\ref{fig1} presents the proposed method. 

SI's first step is to learn an inter-slice synthesis model, which is presented in detail in Fig.~\ref{fig1}(a). Between every two consecutive slices {in the through-plane direction,} the model is used to generate intermediate slices. In this step, the slices of each volume in the through-plane direction are divided into multiple sets in sequence, each with $N+2$ slices, ${\{I_n\}_{n=0}^{N+1}}$, where $N$ denotes the number of intermediate slices between two input slices. Given two input slices $I_{0}$ and $I_{N+1}$, we synthesize the intermediate slices, $\{ \hat{I}_{n}\}_{n=1}^{N}$, which should be as close as possible to the ground-truth intermediate slices $\{I_n\}_{n=1}^{N}$. 

{In particular, the inter-slice synthesis network generates the bidirectional spatial transformations $\hat{F}_{0 \to N+1}$ and $\hat{F}_{N+1 \to 0}$. Then the intermediate spatial transformations $\hat{F}_{n \to 0}$ and $\hat{F}_{n \to N+1}$ can be approximated by combining the bidirectional spatial transformations as follows:}

{
\begin{equation}
\begin{aligned}
\hat{F}_{n \to 0} &= -(1-\frac{n}{N+1})\frac{n}{N+1} \hat{F}_{0 \to N+1} \\
& + \frac{n}{N+1}^{2} \hat{F}_{N+1 \to 0}
\end{aligned}
\end{equation}}

{
\begin{equation}
\begin{aligned}
\hat{F}_{n \to N+1} &= (1-\frac{n}{N+1})^{2} \hat{F}_{0 \to N+1} \\
& - (1-\frac{n}{N+1}) \frac{n}{N+1} \hat{F}_{N+1 \to 0}
\end{aligned}
\end{equation}}

{The intermediate spatial transformations $\hat{F}_{n \to 0}$ and $\hat{F}_{n \to N+1}$ warp $I_{0}$ and $I_{N+1}$, respectively, to synthesize the inter-slice $\hat{I}_n$ as follows:}

\begin{equation}
\begin{aligned}
\hat{I}_{n} = &(1-\frac{n}{N+1}) g(I_{0}, \hat{F}_{n\rightarrow0}) + \frac{n}{N+1} g(I_{N+1}, \hat{F}_{n\rightarrow N+1})
\end{aligned}
\label{eq1}
\end{equation}

where $g(\cdot, \cdot)$ denotes a backward-warping function, which is implemented with bilinear interpolation \cite{b26, b27}.

$\hat{I}_n$ and $I_n$ are further fed into the global discriminator and multitask learning model. The multitask learning model consists of an attention network, an object classifier, and a local discriminator. The object classifier detects whether the input slice contains the target object, whereas the two discriminators determine whether the slices are synthetic or real. The attention network focuses on the useful parts in the slice that help the object classifier and the local discriminator make predictions on their tasks. The object classifier and local discriminator are optimized together, and they share the same attention network, to take advantage of the interacting information between the two tasks. 

The second step of SI is to create synthetic slices and their associated segmentation labels between two consecutive slices {in the through-plane direction} using the learned inter-slice synthesis network. The process of synthesizing intermediate slices and their corresponding labels is presented in Fig.~\ref{fig1}(b). Given two consecutive input slices $I_{0}$ and $I_{1}$, and their corresponding segmentation labels $L_0$ and $L_1$, the learned inter-slice synthesis network generates the intermediate spatial transformations, $\hat{F}_{n \to 0}$ and $\hat{F}_{n \to 1}$, at position $n\in(0,1)$. $I_0$, $I_1$ and $L_0$, $L_1$ are warped by $\hat{F}_{n \to 0}$ and $\hat{F}_{n \to 1}$ to generate the intermediate slice and their corresponding segmentation label, $\hat{I}_n$ and $\hat{L}_n$, as follows:

\begin{equation}
\hat{I}_{n} = (1-n) g(I_{0}, \hat{F}_{n\rightarrow0}) + n g(I_{1}, \hat{F}_{n\rightarrow 1})
\label{eq1_1}
\end{equation}

\begin{equation}
\hat{L}_{n} = (1-n) g(L_{0}, \hat{F}_{n\rightarrow0}) + n g(L_{1}, \hat{F}_{n\rightarrow1})
\label{eq1_2}
\end{equation}

The interpolated slices are used to train the segmentation network in the third step of SI. Fig.~\ref{fig1}(c) depicts the segmentation networks in detail. The synthetic slices and segmentation labels are interpolated into the original volumes {in the through-plane direction} during the training of the 3D segmentation network to convert the 3D volumes into isotropic ones. We then train the 3D segmentation network with the isotropic volumes and the segmentation labels. After the training process, we use the learned inter-slice synthesis network to generate intermediate slices for test samples and interpolate the synthetic slices into the test samples {in the through-plane direction} to make them isotropic. We then remove the output segmentation labels of the synthetic intermediate slices after feeding the interpolated test volumes into the learned 3D segmentation network. The model's performance is evaluated by comparing the remaining parts of the segmentation labels with the ground-truth.

\textbf{The inter-slice synthesis network}: 
We construct the inter-slice synthesis network using the method proposed by Jiang et al. \cite{b28}. The loss function of the inter-slice synthesis network is defined as follows:

\begin{equation}
\begin{aligned}
l =& \lambda_{rec} l_{rec} + \lambda_{per} l_{per} + \lambda_{warp} l_{warp} \\
&+ \lambda_{smooth} l_{smooth} + \lambda_{adv} l_{adv} + \lambda_{tp-smooth} l_{tp-smooth}
\end{aligned}
\label{eq2}
\end{equation}

Equation~\eqref{eq2} is a linear combination of six terms, where a set of coefficients \{ $\lambda_{rec}$, $\lambda_{per}$, $\lambda_{warp}$, $\lambda_{smooth}$, $\lambda_{adv}$, $\lambda_{tp-smooth}$ \} regularizes the contribution of the corresponding term. 

{The first term of \eqref{eq2} is $l_{rec}$. It is the reconstruction loss between the real slices and the synthetic slices. The loss function of $l_{rec}$ is defined as follows:}

{
\begin{equation}
l_{rec} = \frac{1}{N} \sum_{i=1}^{N} \left \| \hat{I}_{n} - I_{n}\right \|_{1}
\end{equation}
}

{The second term of \eqref{eq2} is $l_{per}$. It is the perceptual loss to measure perceptual difference between $\hat{I}_{n}$ and $I_{n}$, which can preserve details of the predictions and make interpolated frames sharper \cite{b28}. The loss function of $l_{per}$ is defined as follows:}

{
\begin{equation}
l_{per} = \frac{1}{N} \sum_{i=1}^{N} \left \| \phi ( \hat{I}_{n}) - \phi ( I_{n})\right \|_{2}
\end{equation}
}

{where $\phi$ means the conv4\_3 features of an ImageNet pre-trained VGG16 model \cite{b28.5}.}

{The third term of \eqref{eq2} is $l_{warp}$. It is the warping loss, which models the quality of the spatial transformation \cite{b28}. The loss function of $l_{warp}$ is defined as follows:}

{
\begin{equation}
\begin{aligned}
l_{warp} & = \left \|I_{0} - g(I_{N+1}, F_{N+1\rightarrow0})\right \|_{1} \\
& + \left \|I_{N+1} - g(I_{0}, F_{0\rightarrow N+1})\right \|_{1} \\
& +\frac{1}{N} \sum_{i=1}^{N} \left \| I_{n} - g(I_{0}, \hat{F}_{0\rightarrow n})\right \|_{1} \\
& + \frac{1}{N} \sum_{i=1}^{N} \left \| I_{n} - g(I_{N+1}, \hat{F}_{1\rightarrow n})\right \|_{1}
\end{aligned}
\end{equation}
}

{The fourth term of \eqref{eq2} is $l_{smooth}$. It is the smoothness loss, which encourages neighboring pixels in the in-plan direction to have similar transformation values \cite{b28}. The loss function of $l_{smooth}$ is defined as follows:}

{
\begin{equation}
l_{smooth} = \left \| \bigtriangledown F_{0\rightarrow N+1} \right \|_{1} + \left \| \bigtriangledown F_{N+1\rightarrow 0} \right \|_{1}
\label{eq11}
\end{equation}
}

{The fifth term of \eqref{eq2} is $l_{adv}$. It is the adversarial loss, which encourage the generator to synthesis image to confuse the discriminator. It can improve the authenticity of the synthetic images. The loss function of $l_{adv}$ is defined as follows:}

{
\begin{equation}
l_{adv} = -\frac{1}{N} \sum_{i=1}^{N}\log LD(Att(\hat{I}_{n})) - \frac{1}{N} \sum_{i=1}^{N}\log GD(\hat{I}_{n})
\end{equation}
}

{where $Att$ means the attention network, $LD$ means the local discriminator, $GD$ means the global discriminator.}

The sixth term of \eqref{eq2} is $l_{tp-smooth}$. It is the smoothness term to encourage adjacent pixels of the interpolated volumes in the through-plane direction to have similar values. The loss function of $l_{tp-smooth}$ is defined as follows:

\begin{equation}
\begin{aligned}
l_{tp-smooth} =& \frac{1}{LW}\sum_{i=1}^{L}\sum_{j=1}^{W}((I_{tp}(i,j-1) - I_{tp}(i,j))^2 \\
&+(I_{tp}(i+1,j) - I_{tp}(i,j))^2)
\end{aligned}
\label{eq3}
\end{equation}

where $I_{tp} \in \{ I_{sagittal}, I_{coronal}\}$ denotes the slices in the through-plane directions: $I_{sagittal}$, the volume slices in the sagittal direction; and $I_{coronal}$, the volume slices in the coronal direction. $I_{tp}(i, j)$ denotes the value in $(i, j)$ of $I_{tp}$, $L$ denotes the length of $I_{LR}$ and $W$ denotes the width of $I_{LR}$.

\textbf{The global discriminator and the multitask learning model}:
An attention network, an object classifier, and a local discriminator are the components of the multitask learning model. While the object classifier detects whether the input slices contain the target objects, the global and local discriminators compete with the inter-slice synthesis network. The attention network can automatically focus on the interacting information between the classification and the distinguishing tasks by training the local discriminator and the object classifier simultaneously. 

The loss function for the global discriminator is defined as follows:

\begin{equation}
l_{global} = - \frac{1}{N} \sum_{n=1}^{N}\log (1 - GD(\hat{I}_{n})) - \frac{1}{N} \sum_{n=1}^{N}\log GD(I_{n})
\label{eq5}
\end{equation}

The loss function for the multitask learning model is defined as follows:

\begin{equation}
\begin{aligned}
l&_{mul} =( - \frac{1}{N} \sum_{t=1}^{N}\log (1 - LD(Att(\hat{I}_{n}))) \\
&- \frac{1}{N} \sum_{n=1}^{N}\log LD(Att(I_{n}))) \\
&+ (-\frac{1}{N} \sum_{n=1}^{N} \hat{Y}_{n} \log OC(Att(\hat{I}_{n})) \\
&- \frac{1}{N} \sum_{n=1}^{N} Y_{n} \log OC(Att(I_{n})))
\end{aligned}
\label{eq6}
\end{equation}

where $OC$ denotes the object classifier network. $Y_{i} \in \{0,1\}$ and $\hat{Y}_{i} \in \{0,1\}$ denote whether the real and synthetic slices contain the target object. $Y_{i} = 1$ and $\hat{Y}_{i} = 1$ if the input original and synthetic slices contain the target object, and $Y_{i} = 0$ and $\hat{Y}_{i} = 0$ otherwise. 

\begin{figure*}
\centerline{\includegraphics[scale=0.4]{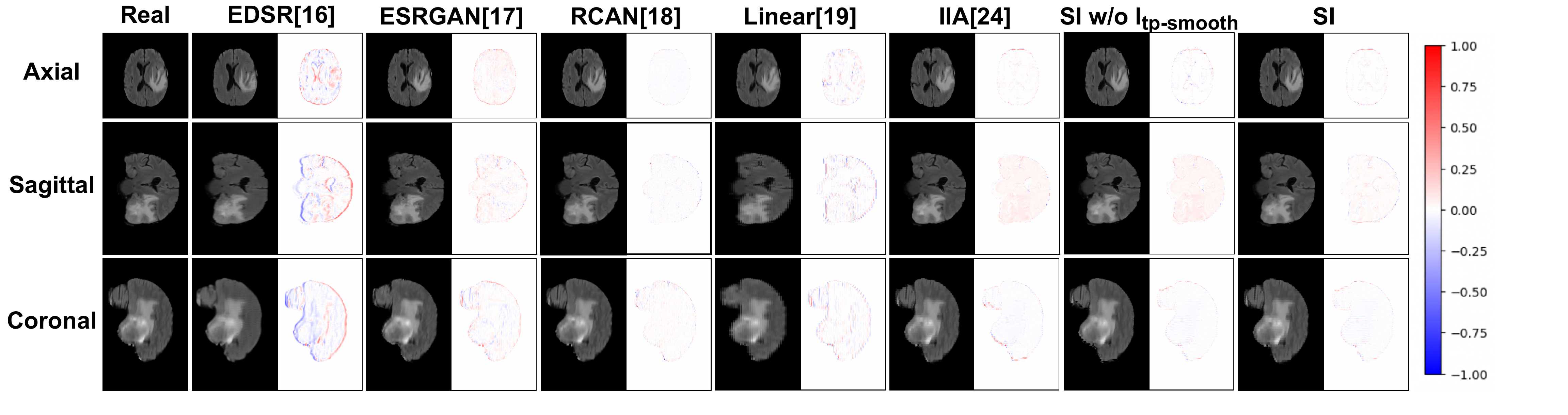}}
\caption{Synthetic slices of different methods in BTS from different directions: the in-plane (axial view), and through-plane (sagittal and coronal views) directions. The difference maps are provided to the right of the results for better visualization.} 
\label{fig4_5}
\end{figure*}

\begin{figure*}
  \centerline{\includegraphics[scale=0.5]{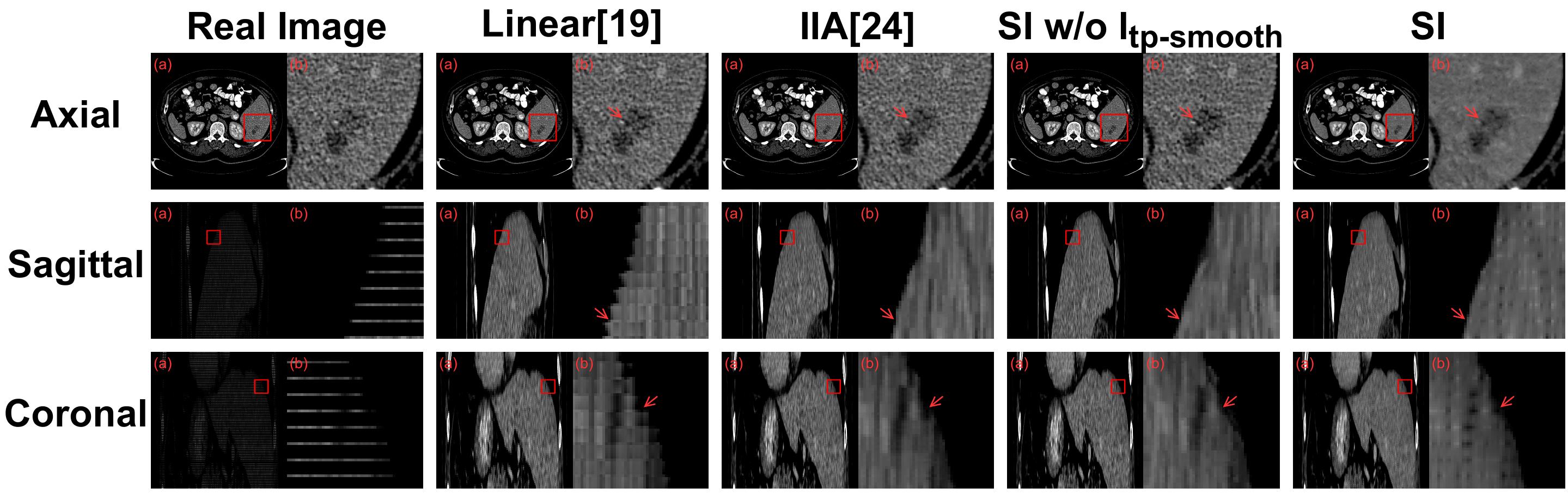}}
    \caption{Synthetic slices of different methods in LTS from different directions: the in-plane (axial view), and through-plane (sagittal and coronal views) directions.}
  \label{fig2}
\end{figure*}

\section{Experiment}
\label{s4}
We provide both quantitative and qualitative performance evaluations for SI on three 3D medical imaging datasets.

\subsection{Datasets}
Brain tumors segmentation (BTS) in Medical Segmentation Decathlon \cite{b29} is the first dataset in our experiment. All scans in BTS are co-registered to a reference atlas space using the SRI24 brain structure template \cite{b29.11}, resampled to isotropic voxel resolution of $1 mm^{3}$, and skull-stripped using various methods followed by manual refinements. We segment the whole tumors in our experiment by selecting 100 FLAIR modal MRI data in this dataset. We use training data from 56 scans, validation data from 14 scans, and test data from 30 scans. 
The second dataset is liver tumor segmentation (LTS) in Medical Segmentation Decathlon \cite{b29}. The LTS slices are generated by a variety of different scanning devices with intra-slice and inter-slice distances ranging from 0.5 to 1mm and 0.45 to 6.0mm, respectively. A total of 131 portal venous phase computed tomography (CT) scans with two annotated objects (liver and tumor) are selected. We use 74, 18, and 39 scans as training, validation, and test data, respectively. 
Prostate segmentation (PS) in Medical Segmentation Decathlon \cite{b29} is the third dataset. PS includes 32 transverse T2-weighted scans with two annotated objects (prostate peripheral zone and the transition zone), each with voxel resolution $0.6 \times 0.6 \times 4 mm^{3}$. We use 17, 6, and 9 scans as training, validation, and test data, respectively.

{
\subsection{Evaluation metrics}
\textbf{Dice score}: Dice score \cite{b29.1} quantifies the overlap between two segmentation labels. The formulation of the Dice score is shown as follows:}

{
\begin{equation}
Dice(L, \hat{L}) = 2\times (\frac{|L\cap \hat{L}|}{|L| + |\hat{L}|}) \times 100 \%
\end{equation}
}

{where $L$ denotes the ground truth of the real image and $\hat{L}$ denotes the predicted segmentation label. If the Dice score is 0, the two labels have no overlap. With the Dice score increasing, the two labels have more overlap. When the Dice score is 1, the two labels have completely overlap. A better model will have a higher Dice score.}

{\textbf{Relative absolute volume difference}: The relative absolute volume difference (RAVD) \cite{b30} reveals if a method tends to over- or undersegment. The formulation of RAVD is defined as follows:}

{
\begin{equation}
RAVD(L, \hat{L}) = (\frac{|\hat{L}| - |L|}{|L|}) \times 100 \%
\end{equation}
}

{A value of 0 means both volumes are identical. A better model will have a lower RAVD.}

{\textbf{Average symmetric surface distance}: Average symmetric surface distance (ASSD) \cite{b30} is given in millimeters and based on the surface voxels of two segmentations $L$ and $\hat{L}$. The formulation of ASSD is defined as follows:}

{
\begin{equation}
\begin{aligned}
AS&SD(L, \hat{L}) = \frac{1}{|S(L)| + |S(\hat{L})|} \\
&(\sum_{l \in S(L)}  \min_{\hat{l} \in S(\hat{L})} \| l - \hat{l} \|_{2} + \sum_{\hat{l} \in S(\hat{L})}  \min_{l \in S(L)} \| \hat{l} - l \|_{2} )
\end{aligned}
\end{equation}
}

{where $S(\cdot)$ denote the set of surface voxels of volumes. For each surface voxel of $L$, the Euclidean distance to the closest surface voxel of $\hat{L}$ is calculated. In order to provide symmetry, the same process is applied from the surface voxels of $\hat{L}$ to $L$. ASSD is then defined as the average of all distances, which is 0 for a perfect segmentation. A better model will have a lower ASSD.}

{\textbf{Maximum symmetric surface distance}: Maximum symmetric surface distance (MSSD) \cite{b30} is given in millimeter and based on the surface voxels of two segmentations $L$ and $\hat{L}$. The formulation of MSSD is defined as follows:}

{
\begin{equation}
\begin{aligned}
MSSD(L, \hat{L}) &= \max \{ \max_{l \in S(L)} \min_{\hat{l} \in S(\hat{L})} \| l - \hat{l} \|_{2}, \\
& \max_{\hat{l} \in S(\hat{L})} \min_{l \in S(L)} \| \hat{l} - l \|_{2} \}
\end{aligned}
\end{equation}
}

{Different from ASSD, surface voxels of MSSD are determined using Euclidean distances, and the maximum value yields MSSD. For a perfect segmentation MSSD is 0. A better model will have a lower MSSD.}

\begin{figure*}
  \centerline{\includegraphics[scale=0.5]{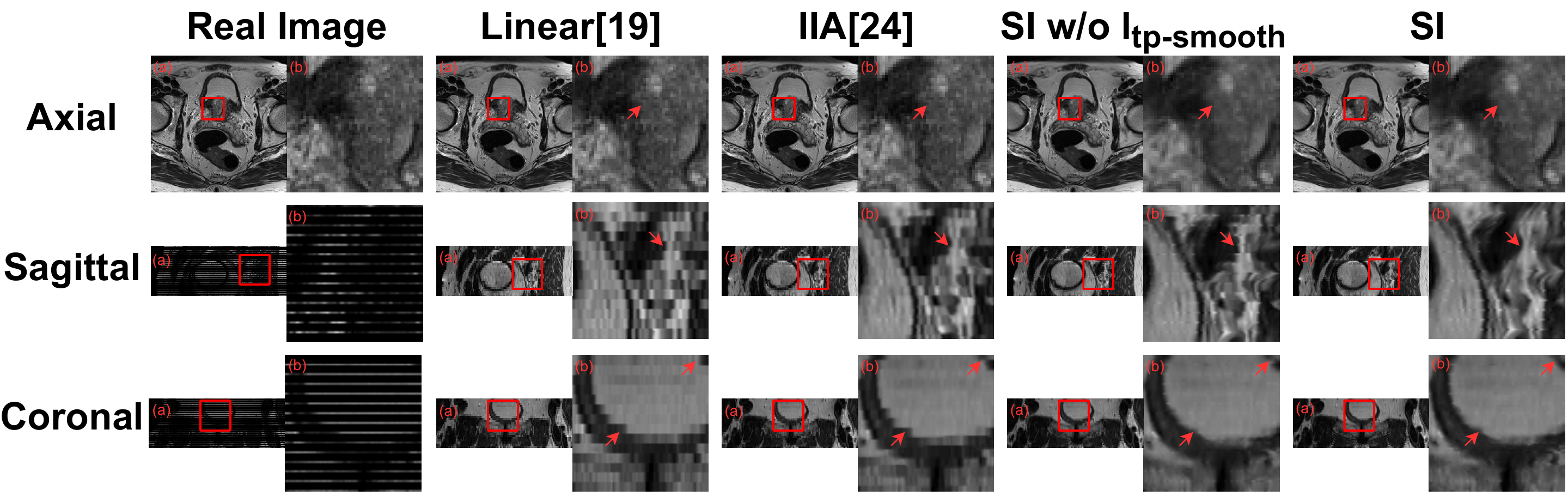}}
    \caption{Synthetic slices of different methods in PS from different directions: the in-plane (axial view), and through-plane (sagittal and coronal views) directions.}
  \label{fig3}
\end{figure*}

\subsection{Implementation}
To implement SI, we divide the training data into multiple sets in sequence, each with $N+2$ slices. We will discuss the setting of $N$ in Section~\ref{sub1}. The first four hyper-parameters of \eqref{eq2} are determined according to \cite{b13}. Furthermore, we use five-fold cross-validation to select $\lambda_{adv}$ and $\lambda_{tp-smooth}$. $\lambda_{adv}$ is set to $0.050$ and $\lambda_{tp-smooth}$ is set to $0.467$. As presented in Fig.~\ref{fig1}(a), we optimize the attention network, object classifier, and local discriminator together. Two fully connected layers comprise the object classifier, whereas, three convolutional layers, a fully connected layer, and a sigmoid function comprise the global and local discriminators. Two branches make up the attention network. The features of the slices are extracted by one convolutional layer, and the corresponding attention masks are generated by two convolutional layers in the other branch. The inter-slice synthesis network is trained in 30 epochs using the basic learning rate of 0.001, and the batch size is set to 2. To optimize all of the networks in the inter-slice synthesis model, we use Adam optimizer \cite{b31}. 

In the experiment of BTS, EDSR\cite{b22_12}, ESRGAN\cite{b22_15}, RCAN\cite{b22_13}, Linear\cite{b18}, and IIA\cite{b13} are compared with our method. The first three methods transform the LR images in the through-plane direction into HR ones and do not generate the corresponding segmentation labels of the synthetic slices. To leverage unlabeled slices, the 2D uncertainty aware self-ensembling mean teacher model\cite{b31_1}, which is a semi-supervised segmentation model, is employed to segment the medical volumes augmented by the first three methods. {The 2D uncertainty aware self-ensembling mean teacher model contains two models, the teacher model and the student model. The backbone of the teacher and student models is U-Net\cite{b14}. For a fair comparison, U-Net is employed to segment the medical volumes augmented by the other methods.}
In the experiments of LTS and PS, Linear\cite{b18}, IIA\cite{b13} are compared with our proposed method. All methods transform the anisotropic medical volumes into isotropic ones and use nnU-Net\cite{b4} to segment the medical volumes.

\subsection{Comparison with SR algorithms}
{In this session, we compare our method’s performance with the baselines on whole tumor segmentation of BTS. To make the data in BTS anisotropic, the $R$-th slices of the volumes in the through-plane direction are removed. $R$ is defined as follows: }

{
\begin{equation}
\begin{aligned}
R = \{ r | r \% 4 \neq 0, 0 \leq r \textless H \}
\end{aligned}
\end{equation}
}

{where $H$ denotes the number of slices of the volumes in the through-plan direction.}

\subsubsection{Visualization performance of synthetic slices}

Synthetic slices of different methods and their difference maps with the original slices in BTS are presented in Fig.~\ref{fig4_5}. The difference maps of EDSR and ESRGAN in Fig.~\ref{fig4_5} indicate that, compared with other methods, the synthetic slices of these methods are significantly different from the original slices (columns 3, 5 in Fig.~\ref{fig4_5}). For Linear, the synthetic slices are less different from the original slices compared with EDSR and ESRGAN. However, the synthetic slices of Linear have aliasing on the edges of the object (columns 9 in Fig.~\ref{fig4_5}). {For RCAN, although it can recover the slices that are similar to the ground-truth, it changes the original slices (column 7 in Fig.~\ref{fig4_5}): thus, it may cause a negative impact on the segmentation and cannot be directly applied to our problem scenario. By contrast, since IIA, SI w/o (without) $l_{tp-smooth}$, and SI interpolate slices into the volumes to make it isotropic, the original slices do not change.} Although IIA and SI w/o $l_{tp-smooth}$ can recover slices that are less different from the original slices on the object's edge, they are unable to recover the texture of the object (columns 11, 13 in Fig.~\ref{fig4_5}). For SI, its performance of recovering on the edges and texture is better than those of IIA and SI w/o $I_{tp-smooth}$, which means that the smoothness function of SI can improve the authenticity of the synthetic slices (columns 15 in Fig.~\ref{fig4_5}).

\subsubsection{Segmentation performance evaluations}

\begin{table}[]
\begin{center}
\caption{Performance on the BTS dataset for whole tumor segmentation, ($\uparrow$ denotes higher is better, whereas $\downarrow$ denotes lower is better). The best results are in \textbf{bold}, and the second-best results are \underline{underlined}.}
\label{t0}
\scalebox{0.85}{
\begin{tabular}{ccccc}
\hline
Method   &  Dice $(\%)\uparrow$&  RAVD $(\%)\downarrow$&  ASSD $(mm)\downarrow$&  MSSD $(mm)\downarrow$          \\ \hline
U-Net\cite{b14} &   82.91          &   18.97          &   2.97          &   13.68        \\
EDSR\cite{b22_12}     &   80.25          &   20.87          &   3.09          &   14.83         \\
ESRGAN\cite{b22_15}   &   81.79          &   19.03          &   3.47          &   16.74         \\
RCAN\cite{b22_13}    &   82.00          &   19.37           &   3.07          &   14.99         \\
Linear\cite{b18}   &   68.69          &   38.33          &   7.75          &   38.06         \\
IIA\cite{b13}   &   85.72          &   38.33          &   2.28          &   8.50         \\
SI w/o  $l_{tp-smooth}$       &   \underline{86.36}         &   \underline{14.71}          &   \textbf{2.14} &   \underline{8.35}          \\
SI    &   \textbf{87.00} &   \textbf{13.16} &    \underline{2.15}         &   \textbf{8.27} \\ \hline
\end{tabular}}
\end{center}
\end{table}

\begin{table*}[]
\begin{center}
\caption{Performance of different methods over two annotated objects, namely, liver and tumor, and the mean scores of two annotated objects in LTS.($\uparrow$ denotes higher is better, whereas $\downarrow$ denotes lower is better). The best results are in \textbf{bold}, and the second-best results are \underline{underlined}.}
\label{t1}
\scalebox{1}{
\begin{tabular}{clllllllllllllll}
\hline
\multicolumn{1}{l}{} & \multicolumn{3}{c}{Dice $(\%)\uparrow$}                                                          &  & \multicolumn{3}{c}{RAVD $(\%)\downarrow$}                                                          &  & \multicolumn{3}{c}{ASSD $(mm)\downarrow$}                                                          &                      & \multicolumn{3}{c}{MSSD $(mm)\downarrow$}                                                          \\ \cline{2-4} \cline{6-8} \cline{10-12} \cline{14-16} 
Method& \multicolumn{1}{c}{Liver} & \multicolumn{1}{c}{Tumor} & \multicolumn{1}{c}{Mean} &  & \multicolumn{1}{c}{Liver} & \multicolumn{1}{c}{Tumor} & \multicolumn{1}{c}{Mean} &  & \multicolumn{1}{c}{Liver} & \multicolumn{1}{c}{Tumor} & \multicolumn{1}{c}{Mean} & \multicolumn{1}{c}{} & \multicolumn{1}{c}{Liver} & \multicolumn{1}{c}{Tumor} & \multicolumn{1}{c}{Mean} \\ \hline

nnU-Net\cite{b4}&\multicolumn{1}{c}{94.11}&\multicolumn{1}{c}{55.81}&\multicolumn{1}{c}{74.96}&&\multicolumn{1}{c}{6.69}&\multicolumn{1}{c}{49.01}&\multicolumn{1}{c}{27.85}&&\multicolumn{1}{c}{3.00}&\multicolumn{1}{c}{13.15}&\multicolumn{1}{c}{8.08}&&\multicolumn{1}{c}{60.53}&\multicolumn{1}{c}{60.48}&\multicolumn{1}{c}{60.51}\\

Linear\cite{b18}&\multicolumn{1}{c}{94.01}&\multicolumn{1}{c}{55.54}&\multicolumn{1}{c}{74.78}&&\multicolumn{1}{c}{7.10}&\multicolumn{1}{c}{46.60}&\multicolumn{1}{c}{26.85}&&\multicolumn{1}{c}{2.60}&\multicolumn{1}{c}{12.91}&\multicolumn{1}{c}{7.76}&&\multicolumn{1}{c}{\underline{51.09}}&\multicolumn{1}{c}{60.19}&\multicolumn{1}{c}{55.64}\\

IIA\cite{b13}&\multicolumn{1}{c}{94.30}&\multicolumn{1}{c}{\underline{56.88}}&\multicolumn{1}{c}{\underline{75.79}}&  &\multicolumn{1}{c}{\textbf{5.76}}&\multicolumn{1}{c}{47.22}&\multicolumn{1}{c}{26.49}& &\multicolumn{1}{c}{\underline{1.97}}&\multicolumn{1}{c}{15.41}&\multicolumn{1}{c}{8.69}&&\multicolumn{1}{c}{\textbf{36.76}}&\multicolumn{1}{c}{68.70}&\multicolumn{1}{c}{53.23}\\

SI w/o  $l_{tp-smooth}$&\multicolumn{1}{c}{\textbf{95.14}}&\multicolumn{1}{c}{56.37}&\multicolumn{1}{c}{75.76}&&\multicolumn{1}{c}{6.49}&\multicolumn{1}{c}{\underline{46.48}}&\multicolumn{1}{c}{\underline{26.48}}&&\multicolumn{1}{c}{2.44}&\multicolumn{1}{c}{\textbf{10.73}}&\multicolumn{1}{c}{\textbf{6.59}}&&\multicolumn{1}{c}{53.43}&\multicolumn{1}{c}{\textbf{49.15}}&\multicolumn{1}{c}{\underline{51.29}}\\

SI&\multicolumn{1}{c}{\underline{94.57}}&\multicolumn{1}{c}{\textbf{57.38}}&\multicolumn{1}{c}{\textbf{75.98}}&&\multicolumn{1}{c}{\underline{6.14}}&\multicolumn{1}{c}{\textbf{44.55}}&\multicolumn{1}{c}{\textbf{25.35}}&&\multicolumn{1}{c}{\textbf{1.95}}&\multicolumn{1}{c}{\underline{12.69}}&\multicolumn{1}{c}{\underline{7.32}}&&\multicolumn{1}{c}{52.09}&\multicolumn{1}{c}{\underline{49.59}}&\multicolumn{1}{c}{\textbf{50.84}}\\ \hline
\end{tabular}}
\end{center}
\end{table*}

\begin{table*}[]
\begin{center}
\caption{Performance over two annotated objects, namely, {prostate peripheral zone (PZ) and the transition zone (TZ)}, and the mean scores of two annotated objects in PS ($\uparrow$ denotes higher is better, whereas $\downarrow$ denotes lower is better). The best results are in {bold}, and the second-best results are \underline{underlined}.}
\label{t2}
\scalebox{1}{
\begin{tabular}{clllllllllllllll}
\hline
\multicolumn{1}{l}{} & \multicolumn{3}{c}{Dice $(\%)\uparrow$}&  & \multicolumn{3}{c}{RAVD $(\%)\downarrow$}&  & \multicolumn{3}{c}{ASSD $(mm)\downarrow$}&& \multicolumn{3}{c}{MSSD $(mm)\downarrow$}\\ \cline{2-4} \cline{6-8} \cline{10-12} \cline{14-16} 
Method& \multicolumn{1}{c}{PZ} & \multicolumn{1}{c}{TZ} & \multicolumn{1}{c}{Mean} &  & \multicolumn{1}{c}{PZ} & \multicolumn{1}{c}{TZ} & \multicolumn{1}{c}{Mean} &  & \multicolumn{1}{c}{PZ} & \multicolumn{1}{c}{TZ} & \multicolumn{1}{c}{Mean} & \multicolumn{1}{c}{} & \multicolumn{1}{c}{PZ} & \multicolumn{1}{c}{TZ} & \multicolumn{1}{c}{Mean} \\ \hline

nnU-Net\cite{b4}&\multicolumn{1}{c}{87.79}&\multicolumn{1}{c}{83.81}&\multicolumn{1}{c}{85.81}&  &\multicolumn{1}{c}{9.66}&\multicolumn{1}{c}{18.98}&\multicolumn{1}{c}{14.32}&  &\multicolumn{1}{c}{0.57}&\multicolumn{1}{c}{0.68}&\multicolumn{1}{c}{0.63}&                      &\multicolumn{1}{c}{4.39}&\multicolumn{1}{c}{\textbf{5.92}}&\multicolumn{1}{c}{5.15}\\

Linear\cite{b18}&\multicolumn{1}{c}{89.78}&\multicolumn{1}{c}{84.74}&\multicolumn{1}{c}{87.26}&  &\multicolumn{1}{c}{8.39}&\multicolumn{1}{c}{17.29}&\multicolumn{1}{c}{12.84}&  &\multicolumn{1}{c}{0.49}&\multicolumn{1}{c}{0.64}&\multicolumn{1}{c}{0.56}& &\multicolumn{1}{c}{3.99}&\multicolumn{1}{c}{6.12}&\multicolumn{1}{c}{5.06}\\

IIA\cite{b13}&\multicolumn{1}{c}{\underline{89.97}}&\multicolumn{1}{c}{85.72}&\multicolumn{1}{c}{87.84}&  &\multicolumn{1}{c}{\underline{7.49}}&\multicolumn{1}{c}{\underline{15.86}}&\multicolumn{1}{c}{11.68}&  &\multicolumn{1}{c}{\underline{0.47}}&\multicolumn{1}{c}{\underline{0.59}}&\multicolumn{1}{c}{\underline{0.53}}& &\multicolumn{1}{c}{3.90}&\multicolumn{1}{c}{6.38}&\multicolumn{1}{c}{5.14}\\

SI w/o  $l_{tp-smooth}$&\multicolumn{1}{c}{89.96}&\multicolumn{1}{c}{\underline{86.19}}&\multicolumn{1}{c}{\underline{88.08}}&  &\multicolumn{1}{c}{7.50}&\multicolumn{1}{c}{\textbf{15.63}}&\multicolumn{1}{c}{\underline{11.56}}&  &\multicolumn{1}{c}{\underline{0.47}}&\multicolumn{1}{c}{\textbf{0.58}}&\multicolumn{1}{c}{\textbf{0.52}}& &\multicolumn{1}{c}{\underline{3.76}}&\underline{5.96}&\multicolumn{1}{c}{\textbf{4.86}}\\

SI&\multicolumn{1}{c}{\textbf{90.29}}&\multicolumn{1}{c}{\textbf{86.33}}&\multicolumn{1}{c}{\textbf{88.31}}&  &\multicolumn{1}{c}{\textbf{6.31}}&\multicolumn{1}{c}{16.46}&\multicolumn{1}{c}{\textbf{11.39}}&  &\multicolumn{1}{c}{\textbf{0.46}}&\multicolumn{1}{c}{\textbf{0.58}}&\multicolumn{1}{c}{\textbf{0.52}}& &\multicolumn{1}{c}{\textbf{3.56}}&\multicolumn{1}{c}{6.27}&\multicolumn{1}{c}{\underline{4.91}}\\ \hline
\end{tabular}}
\end{center}
\end{table*}

The segmentation accuracies obtained by different methods are presented in Table \ref{t0}. As can be seen from Table \ref{t0}, all the SR algorithms perform worse than U-Net. Although SR algorithms can recover slices that are less different from the original slices, they are not suitable for helping with segmentation, because the original slices are changed. Moreover, the training of the SR algorithm needs LR/HR pairs; thus, these methods cannot be trained with LR slices only. Hence, the SR algorithm is not suitable for our problem scenario. Linear has the worst performance in the experiments. One possible reason is that the distance of the original slices is large. It's difficult to recover the feature of the missing slices using linear interpolation in the through-plane direction. {By contrast, IIA, SI w/o  $l_{tp-smooth}$, and SI perform better than U-Net. Because all of the three methods generate intermediate slices and labels without changing the original slices, and can provide more labeled training data.} Furthermore, all of the three methods use the local discriminator, which makes the methods pay more attention to the authenticity of the target object. Thus, the segmentation model can better learn the feature of the target object, and ultimately improve the segmentation performance. SI w/o  $l_{tp-smooth}$ performs better than IIA, which means that the multitask learning mechanism is useful for boosting segmentation performance. In Addition, due to the smoothness loss function, SI performs better than SI w/o  $l_{tp-smooth}$ overall. 

\subsection{Comparison with algorithms that transform the data into isotropic data}
In this section, we compare our method's performance with the baselines on 3D segmentation tasks of LTS and PS. 
{For isotropic volume, the intervals of the in-plane and through-plane directions are equal, namely the inter-slice distance of an isotropic volume is equal to its intra-slice distance. To transform the interpolated volumes into isotropic ones, the synthetic slices can be interpolated into the anisotropic volume in the through-plane direction to reduce its intra-slice distance.}
In the experiment, the number of interpolated slices between two consecutive slices is calculated using the following equation:

 \begin{equation}
N_{A} = \left \lfloor \frac{D_{inter}}{D_{intra}} \right \rfloor - 1
\label{eq7}
\end{equation}
where $D_{inter}$ denotes the inter-slice distance of the volume; $D_{intra}$ denotes the intra-slice distance of the volume; and $\lfloor \cdot \rfloor$ denotes a floor function that outputs the greatest integer that is less than or equal to the input. 

\subsubsection{Visualization performance of synthetic slices}

The synthetic slices of different methods in LTS and PS are presented in Fig.~\ref{fig2} and Fig.~\ref{fig3}. The synthesized slice of SI w/o  $l_{tp-smooth}$ has a clearer outline of tumor (red arrow) than Linear and IIA, as shown in the first row of Fig.~\ref{fig2}. The second and third rows of Fig.~\ref{fig2} and Fig.~\ref{fig3} show that SI w/o $l_{tp-smooth}$ can generate volumes with clearer edges than Linear and IIA. The results described above indicate that the multitask learning mechanism helps the model capture more details of the object and generate more realistic slices. The synthesized slices of SI are less noisy than those of the other methods in Fig.~\ref{fig2} and Fig.~\ref{fig3}. This is because the smoothness loss function of SI aims at encouraging neighboring pixels to have similar values. The proposed method can generate slices with more spatial smoothness in the through-plane directions by reducing noise in the synthetic slices using the smoothness loss function. On the contrary, since Linear, IIA, and SI w/o  $l_{tp-smooth}$ do not incorporate slice smoothness in the through-plane direction, their synthetic slices suffer aliasing on the edges of the object (rows 2 and 3 in Fig.~\ref{fig2} and Fig.~\ref{fig3}). 

\begin{table*}[]
{
\begin{center}
\caption{The segmentation over two annotated objects, namely, liver and tumor, and the mean scores of two annotated objects in LTS for the detailed ablation studies of each modules in our framework. ($\uparrow$ denotes higher is better, whereas $\downarrow$ denotes lower is better). The best results are in \textbf{bold}, and the second-best results are \underline{underlined}.}
\label{as1}
\begin{tabular}{clcclccclccclccclccc}
\hline
\multicolumn{1}{c}{No.}&&\multicolumn{2}{c}{Module} &  & \multicolumn{3}{c}{Dice $(\%)\uparrow$}&  & \multicolumn{3}{c}{RAVD $(\%)\downarrow$}&  & \multicolumn{3}{c}{ASSD $(mm)\downarrow$}&& \multicolumn{3}{c}{MSSD $(mm)\downarrow$} \\ \cline{3-4} \cline{6-8} \cline{10-12} \cline{14-16} \cline{18-20} 
&&GD            & L D          &  & Liver   & Tumor  & Mean  &  & Liver   & Tumor  & Mean  &  & Liver   & Tumor  & Mean  &  & Liver   & Tumor  & Mean  \\ \hline
1&&&&  &94.27&56.26&75.27&  &6.49&53.51&30&  &2.79&\underline{12.2}&7.50&  &59.77&55.72&57.75\\
2&&$\checkmark$& &  &94.3&56.27&75.29&  &6.4&54.86&30.63&  &2.62&\textbf{11.07}&\textbf{6.85}&  &\underline{54.43}&53.47&\underline{53.95}\\
3&& &$\checkmark$&  &\underline{94.35}&\textbf{57.51}&\underline{75.93}&  &\textbf{5.98}&\underline{45.5}&\underline{25.74}&  &\underline{2.1}&12.23&\underline{7.17}&  &58.22&\underline{50.75}&54.59\\
4&&$\checkmark$&$\checkmark$&&\multicolumn{1}{c}{\textbf{94.57}}&\multicolumn{1}{c}{\underline{57.38}}&\multicolumn{1}{c}{\textbf{75.98}}&&\multicolumn{1}{c}{\underline{6.14}}&\multicolumn{1}{c}{\textbf{44.55}}&\multicolumn{1}{c}{\textbf{25.35}}&&\multicolumn{1}{c}{\textbf{1.95}}&\multicolumn{1}{c}{12.69}&\multicolumn{1}{c}{7.32}&&\multicolumn{1}{c}{\textbf{52.09}}&\multicolumn{1}{c}{\textbf{49.59}}&\multicolumn{1}{c}{\textbf{50.84}}\\ \hline  
\end{tabular}
\end{center}
}
\end{table*}

\begin{table*}[]
{
\begin{center}
\caption{The segmentation over two annotated objects, namely, prostate peripheral zone (PZ) and the transition zone (TZ), and the mean scores of two annotated objects in PS for the detailed ablation studies of each modules in our framework. ($\uparrow$ denotes higher is better, whereas $\downarrow$ denotes lower is better). The best results are in \textbf{bold}, and the second-best results are \underline{underlined}.}
\label{as2}
\begin{tabular}{clcclccclccclccclccc}
\hline
\multicolumn{1}{c}{No.}&&\multicolumn{2}{c}{Module} &  & \multicolumn{3}{c}{Dice $(\%)\uparrow$}&  & \multicolumn{3}{c}{RAVD $(\%)\downarrow$}&  & \multicolumn{3}{c}{ASSD $(mm)\downarrow$}&& \multicolumn{3}{c}{MSSD $(mm)\downarrow$} \\ \cline{3-4} \cline{6-8} \cline{10-12} \cline{14-16} \cline{18-20}  
&&GD            & LD           &  & PZ   & TZ  & Mean  &  & PZ   & TZ  & Mean  &  & PZ   & TZ  & Mean  &  & PZ   & TZ  & Mean  \\ \hline
1&&&&  &89.94&83.94&86.94&  &7.64&20.32&13.98&  &0.57&0.70&0.64&  &7.76&7.12&7.44\\
2&&$\checkmark$& &  &90.00&84.53&87.27&  &6.67&18.78&12.73&  &\underline{0.49}&\textbf{0.48}&\textbf{0.49}&  &\underline{3.70}&6.78&\underline{5.24}\\
3&& &$\checkmark$&  &\underline{90.16}&\underline{84.95}&\underline{87.56}&  &\underline{6.54}&\underline{18.12}&\underline{12.33}&  &0.52&0.62&0.57&  &3.84&\underline{6.66}&5.25\\
4&&$\checkmark$&$\checkmark$&&\multicolumn{1}{c}{\textbf{90.29}}&\multicolumn{1}{c}{\textbf{86.33}}&\multicolumn{1}{c}{\textbf{88.31}}&  &\multicolumn{1}{c}{\textbf{6.31}}&\multicolumn{1}{c}{\textbf{16.46}}&\multicolumn{1}{c}{\textbf{11.39}}&  &\multicolumn{1}{c}{\textbf{0.46}}&\multicolumn{1}{c}{\underline{0.58}}&\multicolumn{1}{c}{\underline{0.52}}& &\multicolumn{1}{c}{\textbf{3.56}}&\multicolumn{1}{c}{\textbf{6.27}}&\multicolumn{1}{c}{\textbf{4.91}}\\ \hline  
\end{tabular}
\end{center}
}
\end{table*}

\subsubsection{Segmentation performance evaluations}

\begin{figure}
\centerline{\includegraphics[scale=0.6]{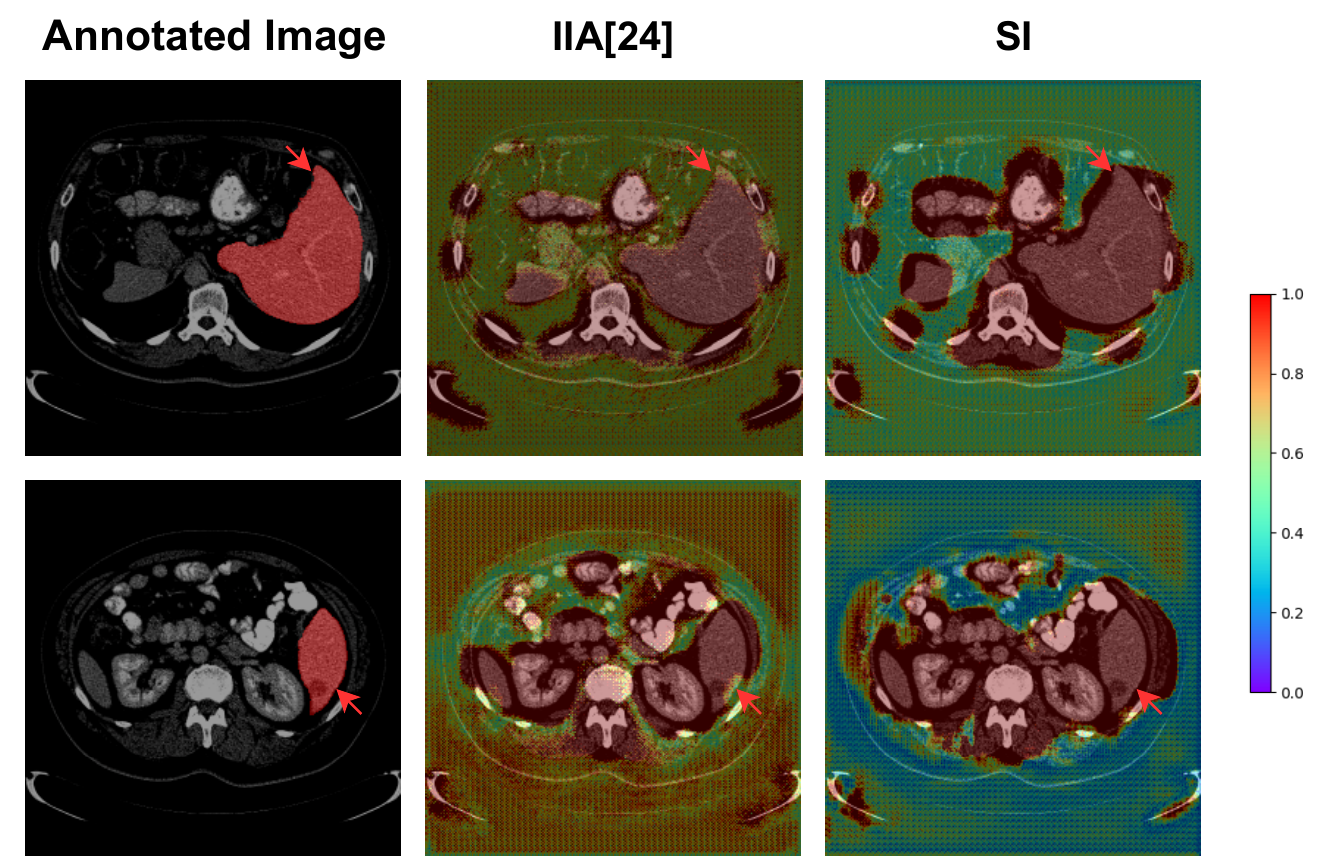}}
\caption{Visualization of the activation maps of the attention networks.} 
\label{fig4}
\end{figure}

Table \ref{t1} and Table \ref{t2} present the 3D segmentation performance of various methods. We also visualize the activation maps extracted by SI's attention networks to demonstrate the effectiveness of the multitask mechanism in SI. Fig.~\ref{fig4} presents the activation maps of two specific slices. Since SI w/o l$_{tp-smooth}$ and SI use the same attention mechanism, we only show the activation maps of SI.

Table \ref{t1} and Table \ref{t2} show that in most cases, nnU-Net and Linear perform worse than IIA, SI w/o  $l_{tp-smooth}$, and SI. Since nnU-Net directly downsamples the HR axes of the volumes to transform the data into isotropic ones, it fails to fully exploit the information of the volumes. Furthermore, Linear directly increases slices resolution, which substantially changes the anatomical structure between consecutive slices. For most cases, SI w/o  $l_{tp-smooth}$ performs better than IIA, suggesting that the multitask learning mechanism helps boosting segmentation performance. Fig.~\ref{fig4} demonstrates that IIA only focuses on the local target object, but SI highlights the whole target object (red arrows in Fig.~\ref{fig4}). The results indicate that the multitask learning model enables the attention network to capture more detailed information of the slices and leads to better segmentation performances. Overall, benefiting from the smoothness loss function, SI achieves the best performance.

{
\section{Discussion}
\label{s4.5}
In this section, we perform a detailed ablation analysis of the network to validate the effect of the local and global discriminators and determine the optimal numbers of intermediate slices. In section \ref{s4.5.1}, we investigate the effect of the global and local discriminators. In section \ref{sub1}, the effect of the number of intermediate slices is studied. }

{
\subsection{Effect of GD and LD}
\label{s4.5.1}
To validate the effect of the local and global discriminators, we conduct detailed ablation studies in LTS and PS. (1) Only the intermediate slice synthesis network is used, without using the two discriminators (global discriminator, $GD$, and local discriminator, $LD$). (2) To highlight the effect of GD, the intermediate slice synthesis network is used to battle with the GD. (3) To highlight the effect of LD, the intermediate slice synthesis network is used to battle with the LD. (4) The intermediate slice synthesis network is used to battle with both the GD and LD.}

{Table~\ref{as1} and Table~\ref{as2} show the segmentation performance under different module configurations. As expected, SI with discriminators outperforms SI without any discriminators. It is worth noting that rows 2 to 4 in Table~\ref{as1} and Table~\ref{as2} have lower ASSD and MSSD than row 1 in Table~\ref{as1} and Table~\ref{as2}. This shows that GD and LD are effective to help SI to generate slices with more realistic edges. Compared with only using GD or LD, using GD and LD simultaneously has the best performance.}
\subsection{Effect of the numbers of intermediate slices}
\label{sub1}
We conduct four segmentation experiments using the inter-slice synthesis network to generate different numbers of intermediate slices between two consecutive slices to investigate the effect of intermediate slice numbers on our method. The results are presented in Table \ref{t3} and Table \ref{t4}. $N_{A}$, the number of automatically interpolated slices, is determined by the original data according to \eqref{eq7}.

\begin{table*}[]
\begin{center}
\caption{Performance of interpolating different numbers of intermediate slices over two annotated objects, namely, liver and tumor, and the mean scores of two annotated objects in LTS ($\uparrow$ denotes higher is better, whereas $\downarrow$ denotes lower is better). $N_{A}$ is the number of interpolated slices determined by the original data. The best results are in \textbf{bold}, and the second-best results are \underline{underlined}.}
\label{t3}
\scalebox{1}{
\begin{tabular}{clllllllllllllll}
\hline
\multicolumn{1}{l}{} & \multicolumn{3}{c}{Dice $(\%)\uparrow$}                                                          &  & \multicolumn{3}{c}{RAVD $(\%)\downarrow$}                                                          &  & \multicolumn{3}{c}{ASSD $(mm)\downarrow$}                                                          &                      & \multicolumn{3}{c}{MSSD $(mm)\downarrow$}                                                          \\ \cline{2-4} \cline{6-8} \cline{10-12} \cline{14-16} 
Interpolated Slices& \multicolumn{1}{c}{Liver} & \multicolumn{1}{c}{Tumor} & \multicolumn{1}{c}{Mean} &  & \multicolumn{1}{c}{Liver} & \multicolumn{1}{c}{Tumor} & \multicolumn{1}{c}{Mean} &  & \multicolumn{1}{c}{Liver} & \multicolumn{1}{c}{Tumor} & \multicolumn{1}{c}{Mean} & \multicolumn{1}{c}{} & \multicolumn{1}{c}{Liver} & \multicolumn{1}{c}{Tumor} & \multicolumn{1}{c}{Mean} \\ \hline

1                    &\multicolumn{1}{c}{94.21}&\multicolumn{1}{c}{\underline{55.15}}&\multicolumn{1}{c}{74.68}&  &\multicolumn{1}{c}{6.63}&\multicolumn{1}{c}{47.72}&\multicolumn{1}{c}{\underline{27.18}}&  &\multicolumn{1}{c}{\underline{2.38}}&\multicolumn{1}{c}{\underline{12.34}}&\multicolumn{1}{c}{7.36}&&\multicolumn{1}{c}{\textbf{47.70}}&\multicolumn{1}{c}{60.11}&\multicolumn{1}{c}{\underline{53.91}}\\

2                    &\multicolumn{1}{c}{90.90}&\multicolumn{1}{c}{48.59}&\multicolumn{1}{c}{69.75}&  &\multicolumn{1}{c}{12.63}&\multicolumn{1}{c}{56.07}&\multicolumn{1}{c}{34.35}&  &\multicolumn{1}{c}{3.60}&\multicolumn{1}{c}{16.25}&\multicolumn{1}{c}{9.93}&&\multicolumn{1}{c}{61.54}&\multicolumn{1}{c}{81.49}&\multicolumn{1}{c}{71.52}\\

3                    &\multicolumn{1}{c}{\textbf{95.11}}&\multicolumn{1}{c}{54.98}&\multicolumn{1}{c}{\underline{75.05}}&  &\multicolumn{1}{c}{\textbf{5.74}}&\multicolumn{1}{c}{\underline{50.51}}&\multicolumn{1}{c}{28.13}&  &\multicolumn{1}{c}{2.44}&\multicolumn{1}{c}{\textbf{10.28}}&\multicolumn{1}{c}{\textbf{6.36}}&&\multicolumn{1}{c}{59.44}&\multicolumn{1}{c}{\underline{54.65}}&\multicolumn{1}{c}{57.05}\\

$N_{A}$            &\multicolumn{1}{c}{\underline{94.57}}&\multicolumn{1}{c}{\textbf{57.38}}&\multicolumn{1}{c}{\textbf{75.98}}&&\multicolumn{1}{c}{\underline{6.14}}&\multicolumn{1}{c}{\textbf{44.55}}&\multicolumn{1}{c}{\textbf{25.35}}&&\multicolumn{1}{c}{\textbf{1.95}}&\multicolumn{1}{c}{12.69}&\multicolumn{1}{c}{\underline{7.32}}&&\multicolumn{1}{c}{\underline{52.09}}&\multicolumn{1}{c}{\textbf{49.59}}&\multicolumn{1}{c}{\textbf{50.84}}\\ \hline
\end{tabular}}
\end{center}
\end{table*}

\begin{table*}[]
\begin{center}
\caption{Performance of interpolating different numbers of intermediate slices over two annotated objects, namely, {prostate peripheral zone (PZ) and the transition zone (TZ)}, and the mean scores of two annotated objects in PS ($\uparrow$ denotes higher is better, whereas $\downarrow$ denotes lower is better). $N_{A}$ is the number of interpolated slices determined by the original data. The best results are in \textbf{bold}, and the second-best results are \underline{underlined}.}
\label{t4}
\scalebox{1}{
\begin{tabular}{clllllllllllllll}
\hline
\multicolumn{1}{l}{} & \multicolumn{3}{c}{Dice $(\%)\uparrow$}                                                          &  & \multicolumn{3}{c}{RAVD $(\%)\downarrow$}                                                          &  & \multicolumn{3}{c}{ASSD $(mm)\downarrow$}                                                          &                      & \multicolumn{3}{c}{MSSD $(mm)\downarrow$}                                                          \\ \cline{2-4} \cline{6-8} \cline{10-12} \cline{14-16} 
Interpolated Slices& \multicolumn{1}{c}{PZ} & \multicolumn{1}{c}{TZ} & \multicolumn{1}{c}{Mean} &  & \multicolumn{1}{c}{PZ} & \multicolumn{1}{c}{TZ} & \multicolumn{1}{c}{Mean} &  & \multicolumn{1}{c}{PZ} & \multicolumn{1}{c}{TZ} & \multicolumn{1}{c}{Mean} & \multicolumn{1}{c}{} & \multicolumn{1}{c}{PZ} & \multicolumn{1}{c}{TZ} & \multicolumn{1}{c}{Mean} \\ \hline

1		&\multicolumn{1}{c}{\underline{89.73}}&\multicolumn{1}{c}{85.42}&\multicolumn{1}{c}{\underline{87.58}}&  &\multicolumn{1}{c}{9.16}&\multicolumn{1}{c}{21.73}&\multicolumn{1}{c}{15.44}&  &\multicolumn{1}{c}{0.93}&\multicolumn{1}{c}{1.11}&\multicolumn{1}{c}{1.02}& &\multicolumn{1}{c}{4.83}&\multicolumn{1}{c}{6.72}&\multicolumn{1}{c}{5.77}\\

2                    &\multicolumn{1}{c}{89.22}&\multicolumn{1}{c}{85.21}&\multicolumn{1}{c}{87.21}&  &\multicolumn{1}{c}{9.51}&\multicolumn{1}{c}{21.60}&\multicolumn{1}{c}{15.55}&  &\multicolumn{1}{c}{0.77}&\multicolumn{1}{c}{0.92}&\multicolumn{1}{c}{0.84}& &\multicolumn{1}{c}{4.39}&\multicolumn{1}{c}{\textbf{5.91}}&\multicolumn{1}{c}{5.15}\\

3                    &\multicolumn{1}{c}{89.49}&\multicolumn{1}{c}{\underline{85.65}}&\multicolumn{1}{c}{87.57}&  &\multicolumn{1}{c}{\underline{8.65}}&\multicolumn{1}{c}{\underline{20.35}}&\multicolumn{1}{c}{\underline{14.50}}&  &\multicolumn{1}{c}{\underline{0.63}}&\multicolumn{1}{c}{\underline{0.77}}&\multicolumn{1}{c}{\underline{0.70}}& &\multicolumn{1}{c}{\underline{4.06}}&\multicolumn{1}{c}{\underline{5.93}}&\multicolumn{1}{c}{\underline{5.00}}\\

$N_{A}$            &\multicolumn{1}{c}{\textbf{90.29}}&\multicolumn{1}{c}{\textbf{86.33}}&\multicolumn{1}{c}{\textbf{88.31}}&  &\multicolumn{1}{c}{\textbf{6.31}}&\multicolumn{1}{c}{\textbf{16.46}}&\multicolumn{1}{c}{\textbf{11.39}}&  &\multicolumn{1}{c}{\textbf{0.46}}&\multicolumn{1}{c}{\textbf{0.58}}&\multicolumn{1}{c}{\textbf{0.52}}& &\multicolumn{1}{c}{\textbf{3.56}}&\multicolumn{1}{c}{6.27}&\multicolumn{1}{c}{\textbf{4.91}}\\ \hline
\end{tabular}}
\end{center}
\end{table*}
 
Table \ref{t3} and Table \ref{t4} show that in most cases, with $N_{A}$ slices added between every two consecutive slices, our method yields the best results (row 4 in Table \ref{t3} and Table \ref{t4}). This is because the distance between two consecutive slices of different volumes is not the same, so if we interpolate a fixed number of slices between two consecutive slices, not all the volumes become isotropic, some may remain anisotropic. By using $N_{A}$ as the number of interpolated slices between two consecutive slices, $SI$ can adaptively transform all the volumes into isotropic ones, which improves the 3D segmentation performance. 

\section{Conclusion}
\label{s5}
In this paper, we propose a novel multitask frame-interpolation-based method for slice imputation whereby the number of slices and corresponding labels is increased between two consecutive slices. SI can generate more realistic 3D medical volumes by evaluating the smoothness of the interpolated 3D medical volumes in the through-plane direction with a smoothness loss function, which improves the accuracy of 3D medical image segmentation. Furthermore, SI can generate as many intermediate slices as needed between two consecutive slices to transform the 3D medical volumes into isotropic ones. The experiments comparing with the baseline methods on BTS, LTS, and PS demonstrate the superior performances of SI. 
The smoothness loss function and the multitask learning model are shown to be effective in ablation experiments. 3D segmentation networks can achieve the best performance if the number of interpolated slices is appropriate to transform the anisotropic 3D medical volumes into isotropic ones. 

{There are still a lot of works to improve the proposed method. The performance of SI may degrade when encountering domain shift, i.e, attempting to apply the learned models on different modalities or organs that have different distributions from the training data. }

{The exploration of the above limitations leads us to future directions. Domain adaptation can be used in our future work to model the shift between datasets of different modalities or organs \cite{b32}. Furthermore, for different modalities, the idea of cross-modality translation can be applied in our model. The idea of CycleGAN\cite{b33} can be added in our model to learn the feature of different modalities \cite{b34}.}





\section*{Acknowledgment}
This work is supported in part by the Natural Science Foundation of Guangdong Province (2020A1515010717), the Fundamental Research Funds for the Central Universities (2019MS073), NSF-1850492 (to R.L.) and NSF-2045804 (to R.L.).

\bibliographystyle{elsarticle-num}
\bibliography{<your-bib-database>}

\begin{thebibliography}{00}



\bibitem{b0} G. Litjens, T. Kooi, B. Ehteshami, A.A.A. Setio, F. Ciompi, M. Ghafoorian, J. A.W.M. van der Laak, B. van Ginneken and C. I. S{\'a}nchez ``A survey on deep learning in medical image analysis,'' in \emph{Medical image analysis,} vol. 42, pp. 60--88, Dec., 2017.

\bibitem{b00} D. D. Ruikar, K. C. Santosh and R. S. Hegadi, ``Automated Fractured Bone Segmentation and Labeling from CT Images," in \emph{Journal of Medical Systems,} vol. 43, No. 3, pp. 1--13, Feb., 2019.

\bibitem{b000} L. Canalini, J. Klein, D. Miller and R. Kikinis, ``Segmentation-based registration of ultrasound volumes for glioma resection in image-guided neurosurgery," in \emph{International Journal of Computer Assisted Radiology and Surgery,} vol. 14, No. 10, pp. 1697--1713, Aug., 2019.

\bibitem{b0.1} Z. Zhou, M. M. R. Siddiquee, N. Tajbakhsh and J. Liang, ``UNet++: A nested U-Net architecture for medical image segmentation," in \emph{Deep Learning in Medical Image Analysis and Multimodal Learning for Clinical Decision Support,} pp. 3--11, Sep., 2018.

\bibitem{b0.2}X. Li, H. Chen, X. Qi, Q. Dou, C. Fu and P. Heng, ``H-DenseUNet: Hybrid Densely Connected UNet for Liver and Tumor Segmentation From CT Volumes," in \emph{IEEE Transactions on Medical Imaging,} vol. 37, no. 12, pp. 2663--2674, Jun., 2018.

\bibitem{b0.3}A. G. Roy, N. Navab and C. Wachinger, ``Concurrent Spatial and Channel ‘Squeeze \& Excitation’ in Fully Convolutional Networks," in \emph{International Conference on Medical Image Computing and Computer-Assisted Intervention,} Granada, Spain, Sep. 16-20, 2018, pp. 421--429.

\bibitem{b1} J. Du, Z. He, L. Wang, A. Gholipour, Z. Zhou, D. Chen and Y. Jia, ``Super-resolution reconstruction of single anisotropic 3D MR images using residual convolutional neural network,'' in \emph{Neurocomputing,} vol. 392, pp. 209--220, Jun., 2020.

\bibitem{b2} J. V. Manj{\'o}n, P. Coup{\'e}, A. Buades, V. Fonov, D. L. Collins, and M. Robles, ``Non-local MRI upsampling,'' in \emph{Medical image analysis,} vol. 14, no. 6, pp. 784--792, Dec., 2010.

\bibitem{b2.5} S. Liu, D. Xu, S. K. Zhou, O. Pauly, S. Grbic, T. Mertelmeier, J. Wicklein, A. Jerebko, W. Cai and D. Comaniciu, ``3D Anisotropic Hybrid Network: Transferring Convolutional Features from 2D Images to 3D Anisotropic Volumes," in \emph{International Conference on Medical Image Computing and Computer-Assisted Intervention,} Granada, Spain, Sep. 16-20, 2018, pp. 851--858.

\bibitem{b3} A. V. Dalca, K. L. Bouman, W. T. Freeman, N. S. Rost, M. R. Sabuncu and P. Golland, ``Medical image imputation from image collections,'' in \emph{IEEE transactions on medical imaging,} vol. 38, no. 2, pp. 504--514, Feb., 2019.

\bibitem{b5} K. Lee, J. Zung, P. Li, V. Jain, and H. S. Seung, ``Superhuman accuracy on the SEMI3D connectomics challenge,'' \emph{arXiv preprint arXiv:1706.00120}, May., 2017.

\bibitem{b4} F. Isensee, J. Petersen, A. Klein, D. Zimmerer, P. F. Jaeger, S. Kohl, J. Wasserthal, G. Kohler, T. Norajitra, S. Wirkert and K. H. Maier-Hein, ``nnU-Net: Self-adapting framework for U-Net-based medical image Segmentation,'' \emph{arXiv preprint arXiv:1809.10486}, Sep., 2018.

\bibitem{b6} Y. Liang, J. Wang, S. Zhou, Y. Gong, N. Zheng, ``Incorporating image priors with deep convolutional neural networks for image super-resolution,'' in \emph{Neurocomputing,} vol. 194, pp. 340--347, Jun., 2016.

\bibitem{b7} B. Lim, S. Son, H. Kim, S. Nah and K.M. Lee, ``Enhanced deep residual networks for single image super-resolution,'' in \emph{2017 IEEE Conference on Computer Vision and Pattern Recognition Workshops (CVPRW),} Honolulu, Hawaii, Jul. 22-25, 2017, pp. 1132--1140.

\bibitem{b8} R. Tanno, D. E. Worrall, A. Ghosh, E. Kaden, S. N. Sotiropoulos, A. Criminisi and D. C. Alexander, ``Bayesian image quality transfer with CNNs: Exploring uncertainty in dMRI super-resolution,'' in \emph{International Conference on Medical Image Computing and Computer-Assisted Intervention,} Quebec City, QC, Canada, Sep. 11-13, 2017, pp. 611--619.




\bibitem{b22_12} B. Lim, S. Son, H. Kim, S. Nah and K. M. Lee, ``Enhanced deep residual networks for single image super-resolution," in \emph{Proceedings of the IEEE conference on computer vision and pattern recognition workshops,} Venice, Italy, Oct. 22-29, 2017, pp. 136--144.

\bibitem{b22_15} X. Wang, K. Yu, S. Wu, J. Gu, Y. Liu, C. Dong, Y. Qiao, C. C. Loy, ``ESRGAN: Enhanced Super-Resolution Generative Adversarial Networks," in \emph{Proceedings of the European Conference on Computer Vision Workshops,} Munich, Germany, Sep. 8-14, 2018.

\bibitem{b22_13} Y. Zhang, K. Li, K. Li, L. Wang, B. Zhong and Y. Fu, ``Image super-resolution using very deep residual channel attention networks," in \emph{Proceedings of the European conference on computer vision,} Munich, Germany, Sep. 8-14, 2018, pp. 286--301.

\bibitem{b18} X. He, S. Yang, G. Li, H. Chang and Y. Yu, ``Non-local context encoder: Robust biomedical image segmentation against adversarial attacks,'' in \emph{Proceedings of the AAAI Conference on Artificial Intelligence,} vol. 33, pp. 865--872, Jul., 2019.

\bibitem{b22_14} I. J. Goodfellow, J. Pouget-Abadie, M. Mirza, B. Xu, D. Warde-Farley, S. Ozair, A. Courville and Y. Bengio, ``Generative Adversarial Networks," in \emph{arXiv preprint arXiv:1406.2661,} Jun., 2014.

\bibitem{b22_14_5} C. Ledig, L. Theis, F. Huszár, J. Caballero, A. Cunningham, A. Acosta, A. Aitken, A. Tejani, J. Totz, Z. Wang, W. Shi, ``Photo-Realistic Single Image Super-Resolution Using a Generative Adversarial Network," in\emph{Proceedings of the IEEE conference on computer vision and pattern recognition,} Honolulu, Hawaii, Jul. 22-25, 2017, pp.4681-4690.

\bibitem{b10} C. Zhao, A. Carass, B. E. Dewey and J. L. Prince, ``Self super-resolution for magnetic resonance images using deep networks,'' in \emph{2018 IEEE 15th International Symposium on Biomedical Imaging (ISBI 2018),} Washington, DC, USA, Apr. 4-7, 2018, pp. 365--368.

\bibitem{b11} C. Zhao, M. Shao, A. Carass, H. Li, B. E. Dewey, L. M. Ellingsen, J. Woo, M. A. Guttman, A. M. Blitz, M. Stone, P. A. Calabresi, H. Halperin and J. L. Prince, ``Applications of a deep learning method for anti-aliasing and super-resolution in MRI,'' in \emph{Magnetic Resonance Imaging,} vol. 64, pp. 132--141, Dec., 2019.


\bibitem{b13} Z. Wu, J. Wei, W. Yuan, J.Wang and T. Tasdizen, ``Inter-slice image augmentation based on frame interpolation for boosting medical image segmentation accuracy,'' in \emph{24th European Conference on Artificial Intelligence,} Santiago de Compostela, Spain, Aug. 29- Sep. 8, 2020, pp. 130--139.

\bibitem{b14} O. Ronneberger, P. Fischer and T. Brox, ``U-Net: Convolutional networks for biomedical image segmentation,'' in \emph{Medical Image Computing and Computer Assisted Intervention,} Munich, Germany, Oct. 5-9, 2015, pp. 234--241.

\bibitem{b15} Y. Qin, K. Kamnitsas, S. Ancha, J. Nanavati, G. Cottrell, A. Criminisi and A. Nori, ``Autofocus layer for semantic segmentation,'' in \emph{Medical Image Computing and Computer Assisted Intervention,} Granada, Spain, Sep. 16-20, 2018, pp. 603--611.

\bibitem{b16} C. Chen, C.Biffi, G. Tarroni, S. Petersen, W. Bai and D. Rueckert, ``Learning shape priors for robust cardiac MR segmentation from multi-view images,'' in \emph{Medical Image Computing and Computer Assisted Intervention,} Shenzhen, China, Oct. 13-17, 2019, pp. 523--531.

\bibitem{b17} C. Chen, D. Qi, H. Chen, J. Qin and P. Heng, ``Synergistic image and feature adaptation: Towards cross-modality domain adaptation for medical image segmentation,'' in \emph{Proceedings of the AAAI Conference on Artificial Intelligence,} vol. 33, pp. 865--872, Jul., 2019.

\bibitem{a17} S. Liu, D. Xu, S. K. Zhou, S. Grbic, W. Cai and D. Comaniciu, ``Anisotropic Hybrid Network for Cross-Dimension Transferable Feature Learning in 3D Medical Images," in \emph{Deep Learning and Convolutional Neural Networks for Medical Imaging and Clinical Informatics,} pp. 199--216, Sep., 2019.

\bibitem{b20} J. Kim, J. K. Lee and K. M. Lee, ``Accurate image super-resolution using very deep convolutional networks," in \emph{Proceedings of the IEEE conference on computer vision and pattern recognition,} Las Vegas, Nevada, Jun. 26- Jul. 1, 2016, pp. 1646--1654.

\bibitem{b21} C. Pham, A. Ducournau, R. Fablet and F. Rousseau, ``Brain MRI super-resolution using deep 3D convolutional networks," in \emph{2017 IEEE 14th International Symposium on Biomedical Imaging (ISBI 2017),} Melbourne, VIC, Australia, Apr. 18-21, 2017, pp. 197--200.

\bibitem{b21.4}C. Shorten, T. M. Khoshgoftaar, ``A survey on Image Data Augmentation for Deep Learning," in \emph{Journal of Big Data,} vol. 6, No. 1, Jul., 2019.

\bibitem{b21.5} N. Tajbakhsh, L. Jyeaseelan, Q. Li, J. N. Chiang, Z. Wu and X. Ding, ``Embracing imperfect datasets: A review of deep learning solutions for medical image segmentation,'' in \emph{Medical image analysis,} vol. 63, Jul., 2020.

\bibitem{21.6} H. R. Roth, C. T. Lee, H. C. Shin, A. Seff, L. Kim, J. Yao, L. Lu and R. M. Summers, ``Anatomy-specific classification of medical images using deep convolutional nets," in \emph{2015 IEEE 12th International Symposium on Biomedical Imaging (ISBI),} New York, NY, USA, Apr. 16-19, 2015, pp. 101-104.

\bibitem{21.7} F. Milletari, N. Navab and SA. Ahmadi, ``V-Net: Fully Convolutional Neural Networks for Volumetric Medical Image Segmentation," in \emph{2016 Fourth International Conference on 3D Vision (3DV),} Stanford, CA, USA, Oct. 25-28, 2016, pp. 565--571.

\bibitem{b23} S. Hauberg, O.Freifeld, A. B. L. Larsen, J. W. Fisher and L. K. Hansen, ``Dreaming more data: Class-dependent distributions over diffeomorphisms for learned data augmentation," in \emph{Artificial Intelligence and Statistics,} pp. 342--350, 2016.

\bibitem{b24} A. Zhao, G. Balakrishnan, F. Durand, J. V. Guttag and A. V. Dalca, ``Data augmentation using learned transformations for one-shot medical image segmentation," in \emph{Proceedings of the IEEE conference on computer vision and pattern recognition,} Long Beach, California, Jun. 16-20, 2019, pp. 8543--8553.

\bibitem{b26} T. Zhou, S. Tulsiani, W. Sun, J. Malik and A. A. Efros, ``View Synthesis by Appearance Flow," in \emph{European conference on computer vision,} Amsterdam, The Netherlands, Oct. 11-14, 2016, pp. 286-301.

\bibitem{b27} Z. Liu, R. A. Yeh, X. Tang, Y. Liu and A. Agarwala, ``Video Frame Synthesis Using Deep Voxel Flow," in \emph{Proceedings of the IEEE International Conference on Computer Vision,} Venice, Italy, Oct. 22-29, 2017, pp. 4463--4471.

\bibitem{b28} H. Jiang, D. Sun, V. Jampani, M. Yang, E. Learned-Miller and J. Kautz, ``Super SloMo: High Quality Estimation of Multiple Intermediate Frames for Video Interpolation," in \emph{Proceedings of the IEEE conference on computer vision and pattern recognition,} Salt Lake City, America, Jun. 18-22, 2018, pp. 9000--9008.

\bibitem{b28.5} K. Simonyan, A. Zisserman, ``Very Deep Convolutional Networks for Large-Scale Image Recognition," in \emph{arXiv preprint arXiv:1409.1556}, Apr., 2015.

\bibitem{b29} A. L. Simpson, M. Antonelli, S. Bakas, M. Bilello, K. Farahani, B. van Ginneken, A. Kopp-Schneider, B. A. Landman, G. Litjens, B. Menze, O. Ronneberger, R. M. Summers, P. Bilic, P. F. Christ, R. K. G. Do, M. Gollub, J. Golia-Pernicka, S. H. Heckers, W. R. Jarnagin, M. K. McHugo, S. Napel, E. Vorontsov, L. Maier-Hein, M. J. Cardoso, ``A large annotated medical image dataset for the development and evaluation of segmentation algorithms," in \emph{arXiv preprint arXiv:1902.09063}, Feb., 2019.

\bibitem{b29.11} S. Bakas, M. Reyes, A. Jakab, S. Bauer, M. Rempfler, A. Crimi, R. T. Shinohara, C. Berger, S. M. Ha, M. Rozycki~\emph{et al.}, ``Identifying the Best Machine Learning Algorithms for Brain Tumor Segmentation, Progression Assessment, and Overall Survival Prediction in the BRATS Challenge," in \emph{arXiv preprint arXiv:1811.02629}, Nov., 2019.

\bibitem{b29.1} L. R. Dice, ``Measures of the Amount of Ecologic Association Between Species," in \emph{Ecology,} vol. 26, no. 3, pp. 297--302, Jul., 1945.

\bibitem{b30} T. Heimann, B. van Ginneken, M. A. Styner, Y. Arzhaeva, V. Aurich, C. Bauer, A. Beck, C. Becker, R. Beichel and G. Bekes~\emph{et al.}, ``Comparison and Evaluation of Methods for Liver Segmentation From CT Datasets," in \emph{IEEE transactions on medical imaging,} vol. 28, no. 8, pp. 1251--1265, Feb., 2009.

\bibitem{b31} D. P. Kingma and J. Ba, ``Adam: A Method for Stochastic Optimization," in \emph{3rd International Conference on Learning Representations,} San Diego, CA, USA, May 7-9, 2015.

\bibitem{b31_1} L. Yu, S. Wang, X. Li, C. W. Fu and P. A. Heng, ``Uncertainty-Aware Self-ensembling Model for Semi-supervised 3D Left Atrium Segmentation," in \emph{Medical Image Computing and Computer Assisted Intervention,} Shenzhen, China, Oct. 13-17, 2019, pp. 605--613.

\bibitem{b32} A. Rozantsev and M. Salzmann, ``Beyond Sharing Weights for Deep Domain Adaptation," in \emph{IEEE TRANSACTIONS ON PATTERN ANALYSIS AND MACHINE INTELLIGENCE,} vol. 41, no. 4, pp. 801-814, Apr., 2019. 

\bibitem{b33} J. Zhu, T. Park, P. Isola, and A. A. Efros, ``Unpaired Image-to-Image Translation using Cycle-Consistent Adversarial Networks," in \emph{Proceedings of the IEEE international conference on computer vision,} Venice, Italy, Oct. 22-29, 2017, pp. 2223-2232.

\bibitem{b34} W. Yuan, J. Wei, J. Wang, Q. Ma, and T. Tasdizen, ``Unified generative adversarial networks for multimodal segmentation from unpaired 3D medical images," in \emph{Medical Image Analysis,} vol. 64, 101731, Aug., 2020.

\end{thebibliography}



\end{document}